# Analysis of the modernization prospects of the WLCG monitoring framework's messaging subsystem

V. Airiian[#]
Dubna State University, Dubna, Russia
Laboratory of Information Technologies, JINR

**ABSTRACT**

The purpose of the project is an analysis of the modernization prospects of the WLCG monitoring framework's messaging subsystem based on Nagios monitoring software and Apache ActiveMQ technologies.

The modernization process demands thorough examination of the existing subsystem to determine the vital upgrade requirements. Thus the work is focused on research of the main underlying technologies, the existing subsystem's structure and revision of its design and used software.

**INTRODUCTION**

The issue of monitoring the state of the processes taking place in the information systems is an important part, and sometimes a prerequisite for their functioning. Worldwide LHC Computing Grid (WLCG) is a complex distributed information system and its proper functioning cannot be confirmed without a thorough monitoring of its state.

To determine the modernization prospects of the WLCG monitoring framework's messaging subsystem the next tasks were stated:

- to study main underlying technologies;
- to analyze the structure of the existing subsystem;
- to consider the requirements for the modernization of the subsystem;
- to revise the logic of the existing subsystem;
- to introduce tools, software and testing framework.

**UNDERLYING TECHNOLOGIES**

One of the major ways to monitor the network infrastructure is to use a special monitoring software. Being the one, Nagios is employed to monitor hosts statuses of WLCG. It is an open-source application for network monitoring that uses the hybrid data collection mode and offers data access through standard OS interfaces [1].

Exchanging data between instances of Nagios requires the use of a message broker which is a program translating a message from the messaging protocol of the sender to the one of the receiver. Apache ActiveMQ is applied as the message broker in WLCG. It is an open-source software written in Java and supporting such protocols as OpenWire, STOMP, AMQP, MQTT [2].

---

[#] wagram@jinr.ru

From all these protocols, STOMP is used to provide network connections between instances of Nagios and Apache ActiveMQ broker. Simple Text Oriented Message Protocol (STOMP) is a simple text-oriented protocol developed for communication with message brokers [3].

**EXISTING SUBSYSTEM STRUCTURE**

The existing messaging subsystem was developed in 2009. The subsystem is fully implemented as Perl-based plugins using GridMon and CPAN set of libraries. Currently the subsystem is severely outdated.

The subsystem is designed to generate notifications on events in the particular subnet, forward all created messages to the broker, retrieve the output message queue from the broker and display system status on the dashboard.

Nagios is the core of the subsystem. Event-tracking is performed by handle_service_check OCSP module. Data exchange is carried out by Nagios plugins running on top of STOMP protocol. Handle_service_check is executed each time the event occurs in the network observed by Nagios. It reads and validates the values of system variables handled by Nagios and containing the parameters of the event.

Further, to include the data into a Message container (managed by Messaging::Message library) it creates a dictionary comprising the parameters formatted using GridMon::MetricOutput library into machine readable form.

After initialization of the Message container instance, it creates an instance of Messaging::Message::Queue DQS and connects to its path specified in the command line. If the path exists, the Message container instance is sent to the queue.

Next, the message is needed to be forwarded to the broker. Nagios SendToMsg plugin reads a Message container from the local message queue and forwards it to the broker input queue via STOMP.

Besides sending messages to the broker receiving messages from the one is provided by the use of Nagios RecvFromQueue plugin. It was designed to read the local message queue of incoming Message containers, convert them to Nagios Passive check string (passive_result = "[%d] PROCESS_SERVICE_CHECK _RESULT;%s;%s;%s;%s\n" %(timestamp, hostname, servicename, status, output)) and write the converted messages into Nagios Command pipe [4].

When Nagios receives a Nagios Passive check string, it updates the status of the specified host, the description of which is contained in the string, according to the standard status codes: OK = 0; WARNING = 1; CRITICAL = 2; UNKNOWN = 3 [5].

**REQUIREMENTS FOR MODERNIZATION**

After examining the subsystem and discussing the project within the development team of CERN IT-SDC-MI department the following main requirements for the modernization were declared:

- migration from Perl to Python programming language providing compatibility with Scientific Linux CERN 5 and 6;
- replacement of CPAN Directory-Queue Messaging::Message::Queue with python-messaging Queue DQS (DIRQ), CPAN Messaging::Message with python-messaging Message;
- replacement of SendToMsg and RecvFromQueue with stompclt forwarding service and introduction of simplevisor controlling service;
- refactoring of the OCSP module and preprocessing module of GridMon::MetricOutput from Perl to Python considering the use of Python libraries and the new message queue;
- development of a new plugin for retrieving incoming messages;
- implementation of secure network connections by SSL (TLS) between Nagios and ActiveMQ.

**REVISION OF THE SUBSYSTEM STRUCTURE**

A revised messaging subsystem structure derives most of the logic from the old one. The core of the subsystem is also Nagios. To replace CPAN libraries, it has been decided to use python-messaging library. OCSP module is called the same name: handle_service_check. The module reads and validates the values of system variables of the event with the use of os standard Python library and method os.environ, then it creates a Python dictionary comprising the parameters to be packed into a python-messaging Message container formatted by Python Metric library, which is a refactored version of Perl GridMon::MetricOutput library. Instead of Perl Messaging::Message::Queue an instance of python-messaging Queue DQS is created and the Message container is sent to the queue.

The forwarding is performed with stompclt service. It reads a Message container from the local message queue into an input module, processes and uploads it to an output module. Depending on they are configured stompclt the instance handles outgoing or incoming events for Nagios. It also supports secure connections via pem certificates by SSL (TLS) protocol [6].

In order to control the permanent operation of stompclt services, simplevisor managing service is used. Bundle of two allows permanent control and handling of message queues, automatic dispatch and reception of messages [7].

In addition to the transfer of messages to the broker stompclt provides receiving messages the help of Nagios mq2nagios plugin acting similarly to RecvFromQueue.

**TOOLS AND SOFTWARE**

In order to write software modules for Nagios upgraded messaging subsystem, Python programming language is utilized. Python interpreter version 2.4.4 has been chosen as the target one being compatible with Scientific Linux CERN 5 and 6. To ease a development process PyCharm Community Edition version 4.0.4 IDE has been selected, as it is compatible with both versions of the operating system [8].

Also the next extra Python libraries are needed: simplejson to serialize and deserialize JSON data and argparse to read command line arguments.

To implement local data queues DIRQ has been chosen as it is already used in the existing subsystem. It provides simple storage of data in plain text format which is great for different clients and programming languages (and also human readable) [9].

For easy interaction between developers and further support CERN remote Git version control system service has been put in use.

In order to distribute a software package containing the messaging subsystem modules in CERN network, YUM manager of RPM packages is considered. Building of RPM packages in CERN infrastructure is carried out by Koji remote service [10].

Unit testing is planned to be performed with the use of PyTest testing tool.

**SUMMARY**

The messaging subsystem of Worldwide LHC Computing Grid monitoring framework was examined and revised regarding modern technologies and instruments. Within the work the following tasks were performed:

- the existing subsystem structure was analyzed;
- technologies and development tools were selected;
- the logic of interaction and data exchanging within the messaging subsystem was revised.

ACKNOWLEDGEMENTS

Author is grateful to Dr. Marian Babik and Dr. Julia Andreeva from CERN IT-SDC-MI department and Prof. Vladimir Korenkov from LIT JINR for valuable discussions, support and supervising.